# Technology & COVID-19: How Reliant is Society on Technology?

Afsana Rahman[1], Dr. Md. Ruhul Amin[2]

The Young Women's Leadership School of Astoria, Queens, NY, USA [1]

Department of Computer and Information Science, Fordham University, Bronx, NY, USA[2]

afsanar@tywls-astoria.org[1], mamin17@fordham.edu[2]

## ABSTRACT

Adolescents, in particular, have experienced great difficulty in adjusting to the abrupt disruption in their day-to-day lives as a result of the COVID-19 pandemic. With virtual education becoming more prominent and social interaction being replaced with electronic devices, adolescents have become accustomed to a technology-reliant lifestyle. The COVID-19 pandemic has not only taken a toll on adolescents' physical well-being through the decrease in physical movement and physical health but has also taken a toll on their mindsets and coping mechanisms. As a result, students have resorted to technology to deal with their mental and physical health issues. Thus, it is important for educators, parents, and teens to better understand the extent to which the use of technology is feasible and whether technology has truly transformed society for the better or for the worse. This article discusses the social implications of technology use for adolescent development and understanding in the context of COVID-19 with attention to the ways in which technology may be advantageous and disadvantageous for teens in the midst of a global pandemic, as well as what the current use of technology implies about societal progression. Through a series of statistical analyses, this article provides an understanding of the ways in which technology can be helpful and harmful in order to help society implement safe technology use amidst the COVID-19 pandemic and beyond.

We have found that technology has impacted society by providing students with a space for self-expression, reflection, and digital communication amidst the pandemic through messaging platforms such as iMessage, Messenger, & Snapchat and Social Media platforms such as Instagram, TikTok, and Facebook. But among technology's benefits lies the consequences of reliance on it. Mental health issues stemming from unrealistic body images, false advertisement, and increased hate crimes on people of color and minoritiy groups have all spread widely with the help and support of technology. Social media and messaging platforms have become a support system for those in fear of COVID-19 while, at the same time, become the root cause of spreading hate, inaccurate representations, and false realities. As technology has been morphed into society's daily tasks and actions, this article may be useful for people of all ages and backgrounds who are interested in understanding the impact of technology on society.

**Public Significance Statement:** Technology has provided teens with a great number of benefits including self-expression, social interaction, access to resources, and virtual education. However, as the use of technology among teens increases, their mental health deteriorates, academic productivity decreases, reliance on false news grows, and society, as a whole, slows progressions. Acknowledging the benefits and detriments associated with the surge of technology use among adolescents is essential for promoting ethical practices of technology for all individuals in society during the COVID-19 pandemic.

**INTRODUCTION**

December 2019 marks the start of an unprecedented, global pandemic in which impact on human behavior is placed on the same scale as the Second World War, the Great Depression, and the 1918 Spanish Flu [1]. In attempts to restrict the COVID-19 pandemic and control widespread affection, the world has resorted to physical and social distancing practices with the support of masks and reduction in physical contact. During these restrictions, however, cooperation was achieved through the replacement of everyday tasks and actions with technological devices. Although the use of technology was promoted immensely in societies worldwide, ordinary people participated in the escalated surge of emerging technologies while the digital divide continued to provide great struggles to those without access to technology [1]. Now, the impact of technological progressions on teens must be evaluated in order to understand the necessary steps that must be taken to encourage righteous use of technology amidst the COVID-19 pandemic for societies beyond the nation and across the world.

A variety of positive outcomes have resulted from the COVID-19 pandemic including but not limited to: supporting friends and family virtually connect, bringing news and information to the public at ease, increasing self-expression, and supporting new learnings. These positives have begun to shape society permanently as social media has become a crucial outlet for self-expression and video-communication services have become a staple in virtual and verbal communication. COVID-19 has shown society how to problem-solve and resort to technology for the completion of everyday tasks. As the world transitions out of the pandemic, technology has provided immense support to guiding society through the obstacles and advancements of the new world.

Despite the beneficial aspects of technology, many issues have arised from the increased use of technology. These issues include but are not limited to: mental health deterioration, increased false advertisement, decreased academic productivity, and lack of attention. These negatives have steered society towards a technology-controlled society where the reliance on technology forces humanity to make irrational choices. COVID-19 has shown society how to rely on technology for the completion of everyday tasks and in return, technology is leaving permanent damage on mental health, way of life, and society as a whole.

Through the use of statistical analysis, we dissected the data of New York City students ages 13-22 to understand how they utilize technology for academics, leisure, and everything in between. The data allowed us to grasp the technological issues and innovations that shape society today and whether it's steering the future of society towards progressions or downfalls. This paper not only provides insight on how technology is utilized today, but also to what extent we can and cannot rely on technology for our actions and decisions.

**RELATED WORK**

As the COVID-19 pandemic has begun, reliance on technology has only increased. As society resorts to technology to take over day-to-day tasks, there have been great changes in academics for students, mental health, and physical wellbeing.

The education system of the United States has changed dramatically due to the introduction of COVID-19. As students began to rely on technology for completion of daily assignments, classwork, and homework, it has resulted in a major impact on their brains. According to The Negative and Positive Impacts of Technology on Education, due to the pandemic a great number of students are being held back from progressing to the following grade level due to their lack of understanding. Schools have long been established to teach students with hands-on learning and individualized guidance in the context of a whole class and because of COVID-19, virtual instruction has posed great consequences on the academic performance of students [2].

As a result of COVID-19, students' physical health has taken a toll as well. Many students rely on school to obtain their exercise and movement for the day. Because of the pandemic, not only have the students seen a significant reduction in their physical activity due to staring at a computer screen all day long, but they've also lost all motivation to get some movement in as well. At school, students are subconsciously engaging in physical activity, whereas, now they're only given the option to move around and get exercise, something many students neglect [2].

The increased use of technology as a result of the COVID-19 pandemic has also led to changes in mental health in students. Due to the pandemic, students have become accustomed to checking their phones for updates on school, family, and entertainment. Because verbal updates are more difficult to obtain, according to Health Policy and Technology, students constantly feel the need to check their phone's notifications. In fact, simply unlocking their phones from instinct gives them reassurance. This constant reassurance has  led students to feel lonely, depressed, anxious, and has also lowered students' self-esteem [3].

As a society, technology's role in civilization must be understood in the broader context of the nation and of the world in order for society to exit the pandemic successfully and ethically. COVID-19's impact on students is important to analyze as multiple factors such as academics, mental health, and physical health have become issues that have all stemmed from the massive surge of technology use. Understanding technology's role in society today through news channels and social media is nearly impossible to understand. We have addressed the drawbacks and innovations surrounding technology in ways that are different from other analyzations. By touching upon the generations that are impacted the most by technology, we've touched upon the aspects of society that are ruled by technology to shed light on how to prevent technology from ruling society further.

**DATA COLLECTION**

This research aims to understand the relationship between technology and students in relation to COVID-19. We used a digital, anonymous survey to collect data from students ages 13-22 who reside in or attend school in one of the five boroughs in New York City: Queens, Bronx, Brooklyn, Manhattan, and Staten Island. Through our survey, our objective was to gain an understanding of the internet applications that students use,  the extent to which social media use and reason of use have been impacted by COVID-19, and the negative and positive effects students claim technology has had on their lives, the extent to which academic use of technology has been impacted by COVID-19, and the validity and reliability of news amidst the COVID-19 pandemic and what this says about a student's understanding of accurate and inaccurate resources.

Our digital survey allowed us to come to terms with students and technology and come to the realization of how technology affects society today and in the future of society. Our survey comprised of four sections: demographics, social media, academics, and news on the internet. The demographics were personal questions to categorize our participants such as geographic location, age, ethnicity, and gender. The social media section were social media-specific questions such as time spent on socials, social media platforms that are used, and the effectiveness of social media during the pandemic. The academic section included questions regarding time spent on academics using technology, the productivity increase or decrease as a result of COVID-19, and how technology has affected our participants' focus on education. Lastly, the section of news in the internet comprises of the types of news that our participants rely on for reliable information.

**METHODS**

We used Chi-Square tests in order to detect correlation between our data. It allowed us to pay attention to the intricate or unexplainable aspects of our research such as how negative effects of technology are more reliant on COVID-19 than the positive effects of technology. Using the Chi-Square test of independence, the age of a student and the internet applications they use are related. There is a correlation between the applications that students use depending on their age. This includes internet applications for leisure use. The applications that students rely on for such use is in correlation with social stigmas and how students cope with such stigmas. Mental health stigmas the coping mechanisms fo societal pressures vary from age group to age group.

Chi-Square test of independence was also used to detect the correlation between the number of hours students spend on academic use of technology and their age group, the older a student is the more hours they spend on technology for academics. Despite the variation in the number of hours that technology is utilized for academics, there is no correlation between the internet applications that students use for applications. Regardless of their age, they all use similar appications to complete assignments or engage in school activities.

# Self-Identified Genders of Surveyed Students

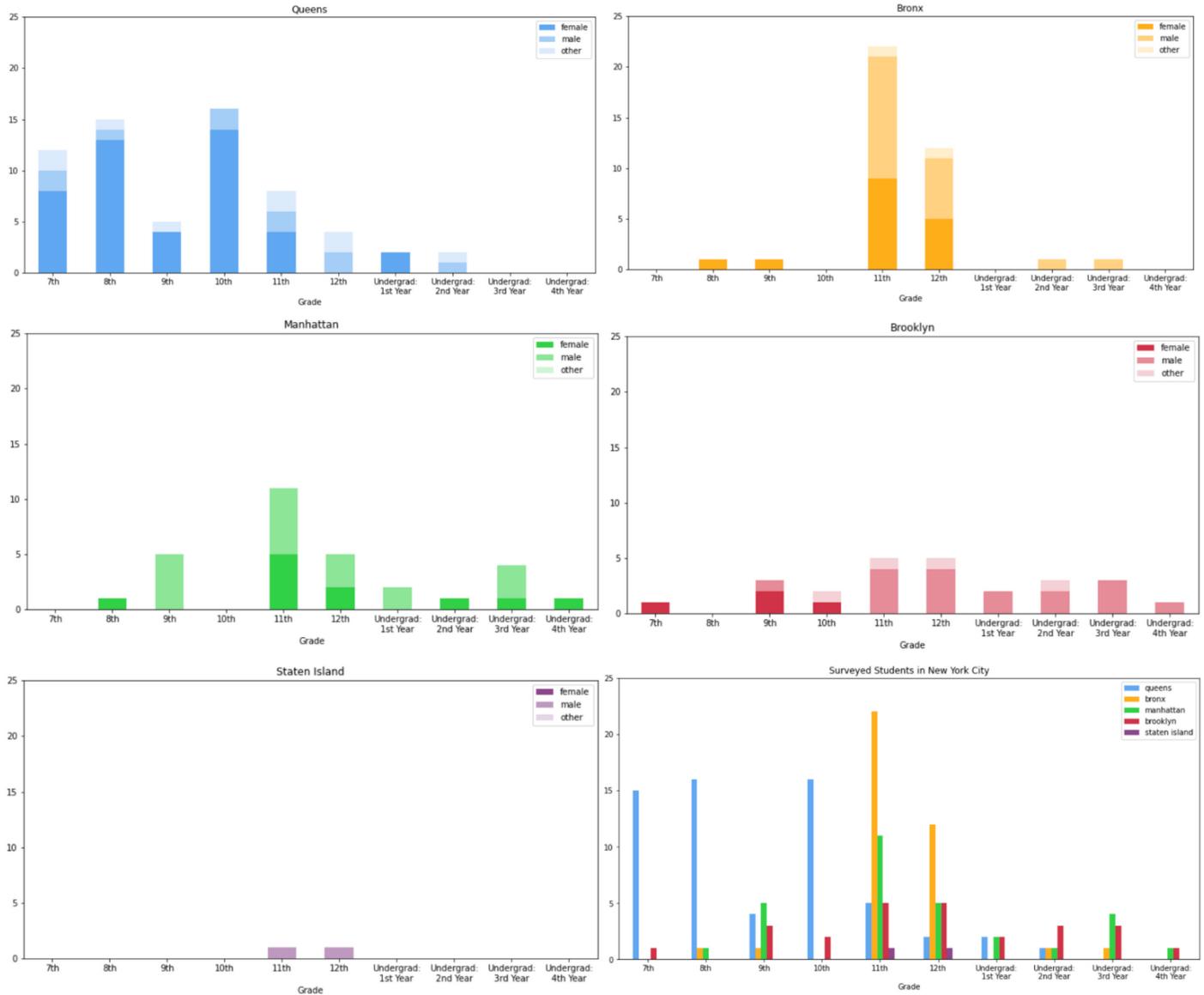

Chi-Square test of independence was also used to come to terms with geography's correlation with technology. Geographical location and the number of hours spent using technology as a whole are relatied, meaning that certain boroughs rely on technology more than others. However, despite the variation in technology utilization, geographical location only impacts the positive effects of technology, not the negative effects. In other words, certain geographical locations benefit more from using technology admist the pandemic than others, but negative effects of technology do not stem from geographical location.

Gender and technology admist the COVID-19 pandemic have no correlation. Using the Chi-Square test of independence, it can be concluded that gender plays no role in the amount of time spent on technology or technology's affect on society's decisions. However, certain genders are more likely to be productive using technology than others.

**UNDERSTANDING**

To add one, certain societal stigmas rise from the use of the technology. Because social media spreads a variety of false information and illogical realities, many societal stigmas stem from the increased leisure use of technology. In other words, the more time spent on social mediator streaming services, the more likely a student is to feel the effects of social stigmas, specifically on mental and physical health. However, social media leads to many positive effects of technology as well. Trustable sources are related to untrustable sources as students' use of technology results in their understanding of which news platforms are reliable and unreliable.

**ç**

In this study, we collected data on students' geographical location in relation to their age. The majority of survey participants reside in either the borough of Queens or the borough of the Bronx. The statistical majority is important in the context of this survey as Queens and Bronx are known to be home to minorities.

While gender doesn't play the most prominent role in the context of the surveyed students' leisure use of technology, it ties to their academic use and females are more likely to claim themselves productive with the support of technology than males.

**LEISURE USE OF TECHNOLOGY**

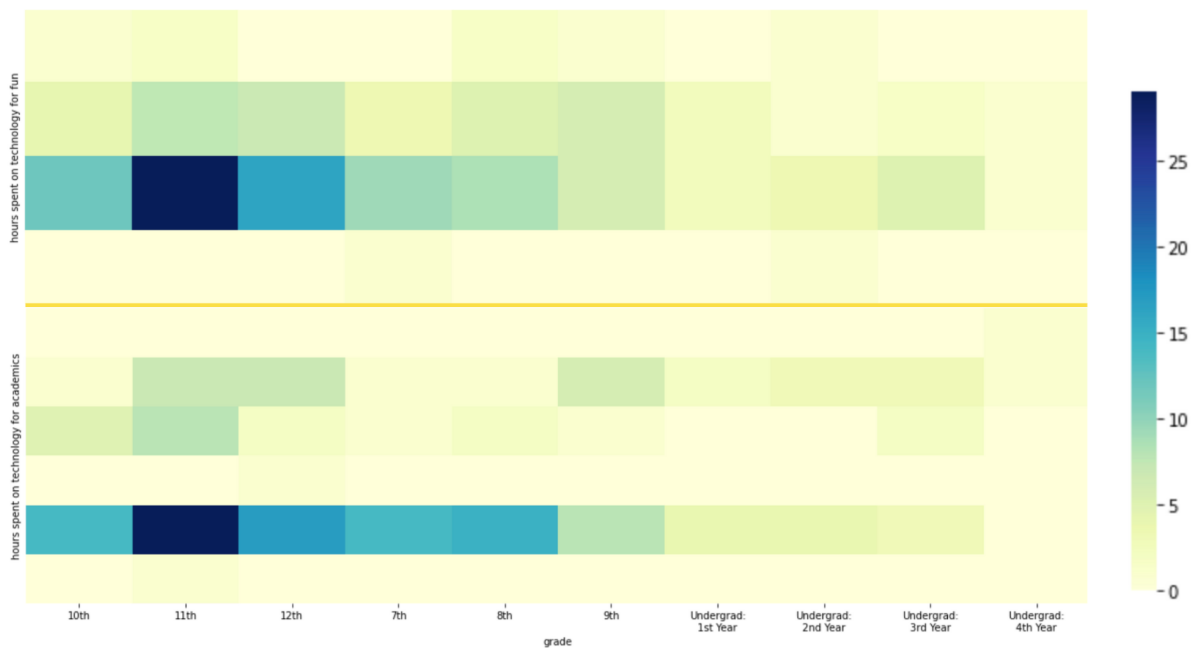

According to the heatmap above, majority of the students spend 4-8 hours on technology for fun. However, despite this extensive use of technology for liesure, the students who have reported 4-8 hours of technolgy use for liesure are the same students who had reported that they spend 6-10 hours on technology for academics. There is a clear correlation between the extensive use of technology for fun and extensive use of technology for academics. Because the United States education system has made it okay for students to engage with technology for acamdeics for long periods of time, students find it okay to use technology for long periods of time for their own pleasure as well. The spike in technology usage across students in New York City has stemmed from the dominoe effect of the surge in technology use as a result of the COVID-19 pandemic. Because businesses, companies,the government, and the education system have resorted to technology over the pandemic, students have had no choice but to feel the same about their personal lives as well.

**SOCIAL STIGMA AND TEENS' COPING MECHANISMS**

Sankey Diagram of societal pressures and coping mechanisms

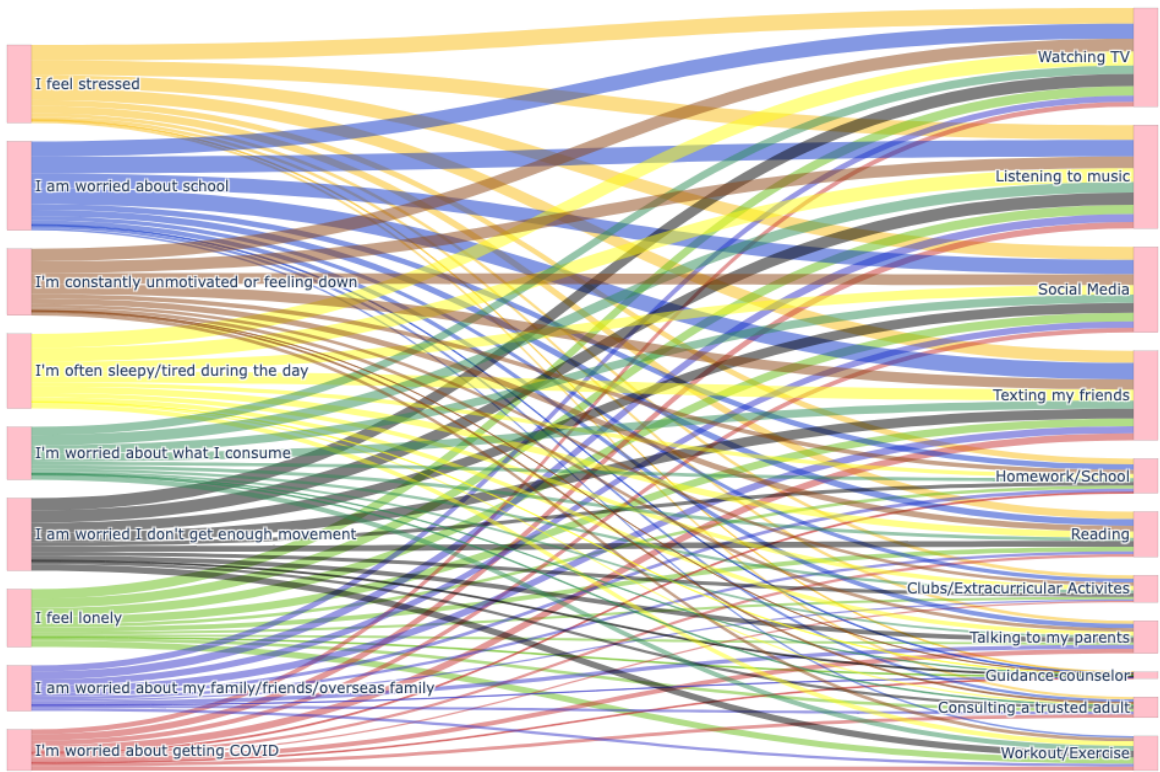

The more students continue to use technology for leisure, the more societal pressures increase. The social stigma of mental health has stemmed mainly from the surge of social media reliance among students. Because of the COVID-19 pandemic, students have had no choice but to communicate with their peers, friends, and family through social media. However, with effective digital communication came the increase in mental health issues as inaccurate information and inaccurate representations flooded social media with false realities. The inaccuracy have social media has affected the mental health of females especially. Students who have been negatively affected by their increased use of social media have reported that they feel unmotivated, stressed, and lonely.

**CONCLUSION**

COVID-19 has resulted in the permanent effects of technology of student life. As more and more issues are crafted in society due to technology, technology is used to cope with such issues, a neverending cycle of using technology to overcome technology. As a society we must overcome technology by not allowing technology to dictate out decisions. The more society relies on technology, the harder it will be to recover from the harsh effects it has. Establishing ethical uses of technology is crucial to societal progression as we move move forward. The more we allow technology to rule our lives, the less control we have on the future of society.

**REFERENCES**

1. Vargo, Deedra, et al. "Digital Technology Use during COVID‑19 Pandemic: A Rapid Review." *Human Behavior and Emerging Technologies*, vol. 3, no. 1, 28 Dec. 2020, pp. 13–24, onlinelibrary.wiley.com/doi/epdf/10.1002/hbe2.242, 10.1002/hbe2.242. Accessed 5 Mar. 2022.
2. Magomedov, I A, et al. "The Negative and Positive Impact of the Pandemic on Education." *Journal of Physics: Conference Series*, vol. 1691, no. 1, 1 Nov. 2020, p. 012134, iopscience.iop.org/article/10.1088/1742-6596/1691/1/012134/meta, 10.1088/1742-6596/1691/1/012134. Accessed 19 Apr. 2022.
3. Ratan, Zubair Ahmed, et al. "Smartphone Overuse: A Hidden Crisis in COVID-19." *Health Policy and Technology*, Jan. 2021, www.ncbi.nlm.nih.gov/pmc/articles/PMC7825859/, 10.1016/j.hlpt.2021.01.002.